\documentclass[11pt,a4paper]{article}
\usepackage{jcappub}
  
\title{The Gravitational Horizon for a Universe with Phantom Energy}
\author[1]{Fulvio~Melia%
\note{John Woodruff Simpson Fellow.}}
\affiliation{Department of Physics, the Applied Math Program, and Steward Observatory \\
              The University of Arizona \\
              Tucson, AZ 85721}
\emailAdd{fmelia@as.arizona.edu}

\abstract{The Universe has a gravitational horizon, coincident with the Hubble sphere, that 
plays an important role in how we interpret the cosmological data. Recently, however, 
its significance as a true horizon has been called into question, even for cosmologies 
with an equation-of-state $w\equiv p/\rho\ge -1$, where $p$ and $\rho$ are the total 
pressure and energy density, respectively. The claim behind this argument is that its
radius $R_{\rm h}$ does not constitute a limit to our observability when the Universe 
contains phantom energy, i.e., when $w<-1$, as if somehow that mitigates the 
relevance of $R_{\rm h}$ to the observations when $w\ge -1$. In this paper, we 
reaffirm the role of $R_{\rm h}$ as the limit to how far we can see sources in the
cosmos, regardless of the Universe's equation of state, and point out that claims 
to the contrary are simply based on an improper interpretation of the null geodesics.
}

\begin{document}
\maketitle
 
 \flushbottom


\section{Introduction}
A newly recognized horizon (dubbed the ``Cosmic Horizon" in [1])  is beginning 
to play a crucial role in our understanding of how the Universe evolves. The distance 
to this horizon is the Universe's gravitational radius, i.e., the radius $R_{\rm h}$ at 
which a proper sphere encloses sufficient mass-energy to turn it into a Schwarzschild 
surface for an observer at the origin of the coordinates (see also [2]). 
Regardless of which specific cosmological model one adopts, $R_{\rm h}$ is always 
given by the expression
\begin{equation}
R_{\rm h}={2GM(R_{\rm h})\over c^2}\;,
\end{equation}
where $M(R_{\rm h})=(4\pi/3) R_{\rm h}^3\,\rho/c^2$, in terms of the total energy 
density $\rho$. Thus, $R_{\rm h}$ is given as $({3c^4/8\pi G\rho})^{1/2}$ which, for a flat
universe, may also be written more simply as $R_{\rm h}(t)=c/H(t)$, where $H(t)$ is the
(time-dependent) Hubble constant.

Although not defined in this fashion, the Hubble radius $c/H(t)$ is therefore clearly
a manifestation of the gravitational horizon $R_{\rm h}$ emerging directly from the 
Robertson-Walker metric written in terms of the proper distance $R=a(t)r$, the expansion
factor $a(t)$, and co-moving radius $r$. What this means, of course, is that the Hubble 
sphere is not merely an empirical artifact of the expanding universe, but actually encloses
a volume centered on the observer containing sufficient mass-energy for its surface to 
function as a {\it static} horizon. That also means that the speed of expansion a distance 
$R_{\rm h}$ away from us must be equal to $c$, just as the speed of matter falling into 
a compact object reaches $c$ at the black hole's horizon---and this is in fact the criterion 
used to define the Hubble radius in the first place. Very importantly, it may also help the
reader to know that since the gravitational radius coincides identically with the Hubble
radius, all of the known equations and constraints (e.g., 2.4 below) satisfied by the latter
must also be satisfied by the former.
 
Since its introduction, however, some confusion regarding the properties of $R_{\rm h}$ 
has crept into the literature, due to a misunderstanding of the role it plays in our observations. 
For example, it has sometimes been suggested  (see, e.g., the earlier paper [3], and 
the more recent work [4]) that sources beyond $R_{\rm h}(t_0)$ are observable 
today (at cosmic time $t_0$), which is certainly not the case. To eliminate misconceptions such as this, some 
effort was made [5] to elaborate upon what the gravitational radius 
$R_{\rm h}$ is---and what it is not. Though first defined in [1], an unrecognized 
form of $R_{\rm h}$ actually appeared in de Sitter's own account of his spacetime 
metric [6]. In reference [5], photon trajectories were calculated for various well-studied 
Friedmann-Robertson-Walker (FRW) cosmologies, including $\Lambda$CDM, demonstrating 
that the null geodesics reaching us at $t_0$ have never crossed $R_{\rm h}(t_0)$. (To be
very precise, what we mean by this is that the proper distance $R_{\gamma}$ of photons
travelling along these null geodesics never attained a value exceeding $R_{\rm h}[t_0]$.)
More generally---and more conclusively---they proved that $R_{\rm h}(t_0)$ is a real limit 
to our observability for any cosmology that does not contain phantom energy [7],
i.e., any cosmology whose equation of state is $p\ge-\rho$, in terms of the total pressure 
$p$ and total energy density $\rho$. (This is a very simple proof and, for completeness, we 
will reproduce it in the discussion below.) 

In spite of this, the issue has been
revisited [8] with the argument that a Universe 
with phantom energy violates this limit, thereby demoting $R_{\rm h}$ from its acknowledgment
as a true horizon. Of course, even if it were true that null geodesics could reach us from beyond
$R_{\rm h}$ when $p<-\rho$, that would simply be highlighting a peculiarity of phantom cosmologies, 
which have other interesting features, such as the acausal transfer of energy. But even in that
scenario, it would have nothing to say about the role played by $R_{\rm h}$ in delimiting what we 
can see in any cosmology that does not contain phantom energy. The purpose of this paper 
is to reiterate what the physical meaning of the Universe's gravitational radius is and to 
demonstrate that, even in cases where $p<-\rho$, no null geodesics reaching us have ever 
crossed the maximum extent of our cosmic horizon.

\section{The Gravitational (Cosmic) Horizon}
Standard cosmology is based on the Friedmann-Robertson-Walker (FRW) metric for a spatially 
homogeneous and isotropic three-dimensional space. In terms of the proper time $t$ measured 
by a comoving observer, and the corresponding radial ($r$) and angular coordinates ($\theta$ 
and $\phi$) in the comoving frame, an interval $ds$ in this metric is written as
\begin{equation}
ds^2=c^2\,dt^2-a^2(t)[dr^2(1-kr^2)^{-1}+r^2(d\theta^2+\sin^2\theta\,d\phi^2)]\;,
\end{equation}
where $a(t)$ is the aforementioned expansion factor and the constant $k$ is $+1$ for a closed
universe, $0$ for a flat, open universe, or $-1$ for an open universe. 

The FRW metric may also be written in terms of the proper distance $R$ [1,9,10], 
and whereas $(ct,r,\theta,\phi)$ describe events in a frame 
``falling'' freely with the cosmic expansion, the second set of coordinates are referenced 
to a particular individual who describes the spacetime relative to the origin at his location.
For a flat universe ($k=0$), it is not difficult to show that
\begin{equation}
ds^2= \Phi\left[c\,dt + \left(\frac{R}{R_{\rm h}} \right)\Phi^{-1} 
dR  \right]^2 - \Phi^{-1}{dR^2}-R^2\,d\Omega^2\;,
\end{equation}
where the function
\begin{equation}
\Phi\equiv 1-\left(\frac{R}{R_{\rm h}} \right)^2
\end{equation}
signals the dependence of the metric on the proximity of the proper radius
$R$ to the gravitational radius $R_{\rm h}$. The exact form of $R_{\rm h}$ 
depends on the composition of the Universe. For example, in de Sitter space 
(which contains only a cosmological constant $\Lambda$), $R_{\rm h}=c/H_0$,
in terms of the (time-invariant) Hubble constant $H_0$. It is equal to $2ct$ in 
a radiation-dominated universe and $3ct/2$ when the Universe contains only 
(visible and dark) matter.

The reason why $R_{\rm h}$ is an essential ingredient of the metric written in 
the form of Equation~(2.2) can be understood in the context of the Birkhoff 
theorem [11] and its corollary.\footnote{We won't reproduce the contextual argument here, 
but instead refer the interested reader to the earlier discussion in [1] and [2].} 
Yet the physical meaning of $R_{\rm h}$ is still elusive to many, possibly 
because of the widely held belief that all horizons must necessarily be asymptotic 
surfaces attained when $t\rightarrow\infty$. The so-called event horizon (see 
[12]) is indeed of this type, representing the ultimate limit to our 
observability to the end of time. However, $R_{\rm h}$ is not in this 
category---nor should it be. Unlike the event horizon, the gravitational 
radius is a time-dependent quantity that increases in value at a rate determined 
by the evolving constituents of the Universe, specifically, the value of the 
equation-of-state parameter $w$, defined by the relation $p=w\rho$. For some 
cosmologies, $R_{\rm h}$ may turn into the event horizon when the cosmic 
time approaches infinity.

It is not difficult to see that the evolution of $R_{\rm h}$ is given by the equation
\begin{equation}
\dot{R}_{\rm h}={3\over 2}(1+w)c\;.
\end{equation}
This is easy to demonstrate from the definition of $R_{\rm h}$ in Equation~(2.1)
and the Friedmann equation. But as we mentioned earlier, it is also known
that the Hubble radius satisfies this equation, so the gravitational radius
must satisfy it as well.
Clearly, $R_{\rm h}$ is constant only for the de Sitter metric, where 
$w=-1$ and therefore $\dot{R}_{\rm h}=0$. For all other values of $w>-1$,
$\dot{R}_{\rm h}>0$. So in $\Lambda$CDM, for example, where the Universe
is currently dominated by a blend of matter and dark energy, $\dot{R}_{\rm h}>0$.
If dark energy were a cosmological constant, however, the Universe would eventually
become de Sitter as the density of matter drops to zero, and we would therefore
expect $R_{\rm h}$ to then asymptotically approach a constant value equal to the
radius of the event horizon in $\Lambda$CDM. What this means, then, is that the 
current location of $R_{\rm h}$ affects what we can observe right now, at time 
$t_0$ since the big bang; it is not---and is not meant to be---an indication of how 
far we will see in our future.

\section{Cosmologies Without Phantom Energy}
We have found it helpful to discuss the properties of null geodesics in FRW 
cosmologies without resorting to conformal diagrams. As 
first shown in [13], it is easier for us to think in terms of familiar quantities (proper 
distances and proper time) that are not always straightforward to interpret 
otherwise. These authors clearly delineated true horizons from apparent
horizons, and extended the definitions, first introduced by [12],
in a clear and pedagogical manner.

Using a similar approach, it was shown in [5] that in all cosmologies without 
phantom energy, all light reaching us today, including that from the recombination 
region associated with the cosmic microwave background (CMB), has traveled a 
{\it net} proper distance of at most $\sim 0.5ct_0$. Because all causally 
connected sources in an expanding universe began in a vanishingly small volume 
as $t\rightarrow 0$, the maximum proper distance from which we receive light 
today must necessarily be finite (and proportional to $ct_0$), since presumably 
there were no pre-existing sources at a non-zero proper distance prior to $t=0$ 
with which we were in causal contact.

The proof that $R_{\rm h}(t)$ is the limit to our observability at any time $t$ is quite straightforward,
and we reproduce it here for completeness. To begin with, we first derive the equation 
governing photon trajectories in a cosmology consistent with the FRW metric (Equation~2.1). 
From the definition of proper distance, we see that
\begin{equation}
\dot{R}=\dot{a}r+a\dot{r}\;.
\end{equation}
But the null condition for geodesics (see, e.g., [13])
applied to Equation~(2.1) yields
\begin{equation}
c\,dt=-a(t){dr\over\sqrt{1-kr^2}}\;,
\end{equation}
where we have assumed propagation of the photon along a radius 
towards the origin. The best indications we have today are
that the universe is flat so, for simplicity, we will assume
$k=0$ in all the calculations described below, and therefore
(for a photon approaching us)
$\dot{r}=-c/a$. Thus, we can write Equation~(3.1) as follows:
\begin{equation}
\dot{R}_\gamma=c\left({R_\gamma\over R_{\rm h}}-1\right)\;,
\end{equation}
where we have added a subscript $\gamma$ to emphasize the
fact that this represents the proper distance of a photon
propagating towards us. Note that in this expression, both
$R_\gamma$ and $R_{\rm h}$ are functions of cosmic time $t$.
The gravitational radius must therefore be calculated according
to Equation~(2.4).

Photons reaching us at $R_\gamma(t_0)=0$ attained a maximum proper distance 
$R_\gamma^{\rm max}$ at time $t_{\rm max}$ which, according to Equation~3.3,
occurred when $\dot{R_\gamma}=0$. At this turning point, $R_\gamma(t_{\rm max})
=R_{\rm h}(t_{\rm max})$. But for all $w\ge-1$, Equation~(2.4) shows that
$\dot{R_{\rm h}}\ge 0$, and therefore $R_{\rm h}(t)\ge R_{\rm h}(t_{\rm max})$
for all $t\ge t_{\rm max}$. It is therefore clear that $R_{\rm h}(t)\ge
R_\gamma(t_{\rm max})\equiv R_\gamma^{\rm max}$ for all $t\ge t_{\rm max}$.
Reference [8] contests this conclusion, but without an
explanation for why this result is incorrect. In the absence of any evidence to the
contrary, one must therefore conclude that $R_{\rm h}(t_0)$ is indeed the limit to our
observability today for any cosmology with $w\ge -1$---and this result stands irrespective of 
whether this limit may be violated by alternative cosmologies with phantom energy. 

\section{Phantom Energy}
The identification of dark energy as a cosmological constant is sometimes called into question 
because of its very low density compared to expectations of the quantum vacuum. Many 
workers have attempted to circumvent these difficulties by proposing alternative forms of 
dark energy, including Quintessence [14,15], which represents 
an evolving canonical scalar field with an inflation-inducing potential, a Chameleon field
(see, e.g., [16,17,18]) in which the
scalar field couples to the baryon energy density and varies from solar system to cosmological 
scales, and modified gravity, arising out of both string motivated, or General Relativity 
modified actions [19,20,21], 
which introduce large length scale corrections modifying the late time evolution of the 
Universe. The actual number of suggested remedies is far greater than this small, 
illustrative sample.

Of these, the most unusual alternative cosmology appears to be the one in which dark energy
has the equation-of-state parameter $w_{\rm de}<-1$ [7,22].
Dubbed ``phantom" energy, only the presence of such a constituent can lead
to a Universe in which the overall equation-of-state parameter $w$ can also be $<-1$, the
situation highlighted in [8] as ``proof" that $R_{\rm h}$ does
not represent a true horizon, even for universes with the equation-of-state parameter
$w\ge -1$, which---as we have just reiterated---is demonstrably incorrect.

Before we consider the null geodesics in this kind of universe, let us first quickly review
the properties of a phantom cosmology. The energy conservation equation in general
relativity tells us that as the Universe expands,
\begin{equation}
\dot\rho=-3H(\rho+p)=-3H\rho(1+w)\;.
\end{equation}
Therefore, if the presence of a phantom energy causes $w<-1$, the overall energy 
density in the Universe grows with time, which seems counter-intuitive because expansion 
should actually deplete the energy supply.\footnote{Note that even in de Sitter, one
often has difficulty justifying why $\rho$ remains constant during the expansion, 
though arguments based on properties of the false vacuum can gain some
traction. But phantom energy is much worse, since even a false vacuum presumably
has a limited energy supply.} Related to this is the fact that the only 
possible way to obtain $w_{\rm de}<-1$ is by violating the so-called dominant energy 
condition (DEC), which for a perfect cosmological fluid may be stated as
\begin{equation}
\rho_{\rm de}\ge |p_{\rm de}|
\end{equation}
where, in obvious notation, $\rho_{\rm de}$ and $p_{\rm de}$ refer to the density
and pressure, respectively, of the phantom energy field. 

\begin{figure}
\begin{center}
\includegraphics[width=0.9\linewidth]{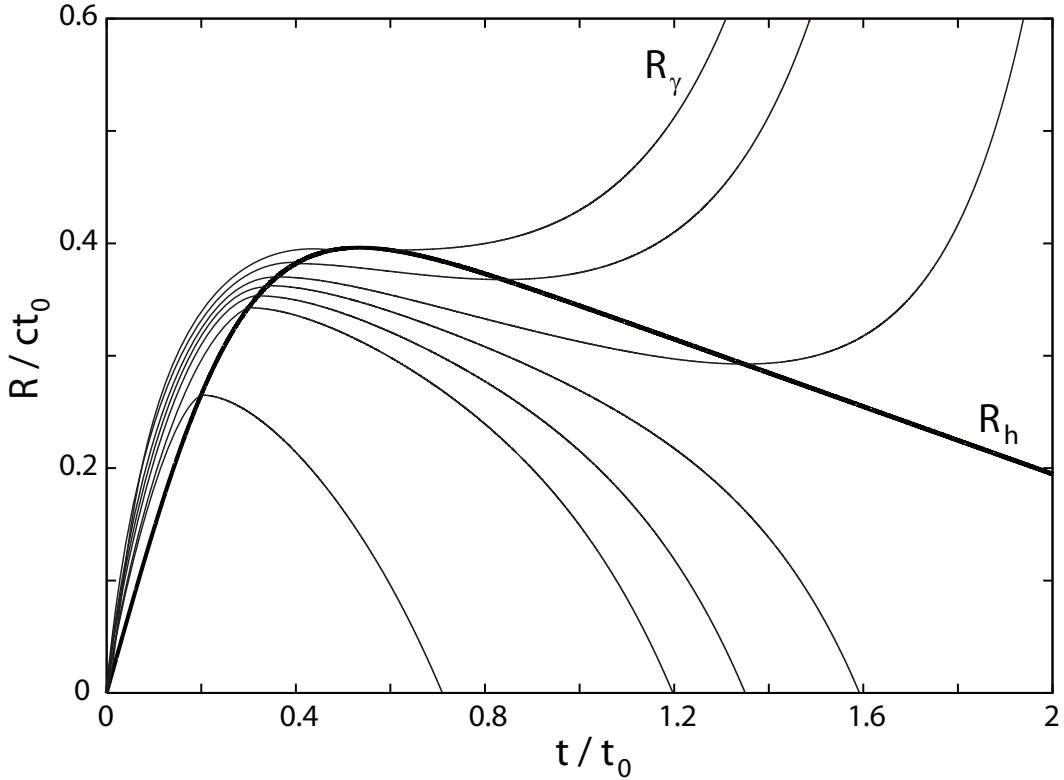}
\end{center}
\caption{Proper distance $R$ in units of $ct_0$ versus $t$ in units of $t_0$, where
$t_0=13.7$ Gyr is the current age of the Universe within the context of the standard
model. The thick curve labeled $R_{\rm h}$ is the gravitational radius of a $\Lambda$CDM-like
universe with dark energy replaced by a phantom field with equation-of-state parameter
$w_{\rm de}=-1.1$. The thin curves represent null geodesics, all of which approach 
$R_\gamma(0)=0$ as $t\rightarrow 0$.  In each case, $R_\gamma$ is the time-dependent
proper radius of photons travelling along a given null geodesic, distinguished by the (chosen) time
$t_{\rm max}$ at which $\dot{R}_\gamma=0$ (see Equation~3.3). The most important feature 
of these curves is that none of those actually reaching us at $R_\gamma(t)=0$ (for $t>0$) 
ever attain a proper distance greater than the maximum extent of our cosmic horizon.
\label{fig1}}
\end{figure}

The physical motivation for the DEC is to prevent instability of the vacuum or
propagation of energy outside the light cone, a situation arising from the
fact that the sound speed in such a fluid is greater than $c$. In a finite time after
the big bang, both the expansion factor and energy density of the phantom field
blow up, leading to a phenomenon known as the ``big rip" [22].
Therefore, a cosmology with permanent phantom energy appears to have very little resemblance
to what we actually see in nature. Under most optimistic assumptions [23], the 
instability timescale in such a scenario can be greater than the current age of the Universe, 
but only in a highly contrived manner. 

Nonetheless, there are situations involving phantom energy that do not lead to a big rip,
and that could also be consistent with the current data. It is therefore important to
understand the behavior of $R_{\rm h}$, and whether it limits the size of the visible 
universe, even when $w<-1$. An interesting class of such cosmologies is the
Scalar-Tensor Models  [24,25], in which the combination of two functions
of the field in the Lagrangian, derived from some underlying theory (such as 
brane cosmology), provides enough flexibility for the Universe to experience
a temporary phase of phantom energy, while still crossing the phantom divide
line (i.e., $p=-\rho$) some time in the future, thereby preventing the otherwise
unavoidable big rip. 

What would null geodesics look like in a cosmology with phantom energy? 
Figure~1 shows $R_{\rm h}(t)$ as a function of cosmic
time $t$ (thick black curve) for a $\Lambda$CDM-like Universe in which the dark energy
component is replaced by a phantom field with $w_{\rm de}=-1.1$. Blending this
energy with matter and radiation at early times, such a Universe's overall
equation-of-state parameter $w$ is still $>0$ until $t\sim 0.5 t_0$, after which 
$w_{\rm de}$ takes over and $w$ drops to negative values. (Note that simply for 
convenience, we scale all of our quantities here to the current age of the Universe, 
$t_0\approx 13.7$ Gyr, in the standard model.) It is easy to see from Equation~2.4 
that $\dot{R}_{\rm h}<0$ subsequent to this transition, and therefore $R_{\rm h}
\rightarrow 0$ at a time corresponding to the ``big rip" when both $a(t)$ and $\rho(t)$ 
diverge to infinity. 

Several of the null geodesics available in such a cosmology are indicated by the thin curves 
in this figure. In each case, $R_\gamma$ is the proper radius of photons travelling along
a given null geodesic, distinguished by the value of $t_{\rm max}$ at which $\dot{R}_\gamma=0$
in Equation~(3.3). In other words, for any given cosmic time $t$, sources may be distributed
throughout the Universe, and therefore the null geodesics passing through those sources
all follow different trajectories $R_\gamma(t)$ bringing them to the origin ($R_\gamma=0$),
where the observer is located. But for each observation time $t_0$, only one null geodesic
connects the observer to all the sources emitting light along its trajectory shown in this figure.

Each of these curves, the gravitational horizon and every null geodesic, approaches 0 as
$t\rightarrow 0$, as one would expect in a universe where no source existed a finite
distance from us prior to the big bang. In addition, in order for us to actually see the light
propagating along a given geodesic, that curve must reach $R_\gamma(t)=0$ for
some time $t>0$. Therefore, according to Equation~3.3, the time at which $R_\gamma=
R_{\rm h}$ corresponds to a turning point in $R_\gamma(t)$, and is in fact a maximum
for the proper distance of every geodesic actually reaching us (just as it was for the
non-phantom cosmologies we considered earlier). 

Reference [8] argued that the existence of geodesics that reverse paths and 
re-cross $R_{\rm h}$ violate its function as a horizon. But what they fail to recognize 
is that every null geodesic that possesses a second turning point  (where again $\dot{R}_\gamma=0$, 
corresponding to another time at which $R_\gamma=R_{\rm h}$ in Equation~3.3), diverges to infinity.
It never again comes back to the origin where we are located. In other words, it is
essential to remember the rather simple notion that in order for us to see light
reaching us from distant sources, the null geodesics must actually 
end at $R_\gamma=0$, not at infinity. And the most important feature illustrated
by this diagram is that none of the geodesics reaching us at $R_\gamma=0$ ever
attained a proper distance greater than the maximum extent of our cosmic (or
gravitational) horizon.    

\section{Concluding Remarks}
We have reaffirmed the physical significance of the Universe's gravitational
horizon, which also defines the size of the Hubble sphere. For any non-phantom
cosmology, we have proven that its current value, $R_{\rm h}(t_0)$, is the maximum 
proper distance to any source producing light in the past that is reaching us today. 
In a Universe with unbounded energy growth, $R_{\rm h}$ eventually shrinks 
to zero when the Universe ends its life in a ``big rip." But even in this case, 
we have demonstrated that no null geodesic reaching us
at $R_\gamma=0$ will have ever attained a proper distance greater than the maximum
extent of $R_{\rm h}$. When $w\ge -1$, $R_{\rm h}$ grows indefinitely, and
therefore its maximum extent simply happens to be its value today, $R_{\rm h}(t_0)$. 
When $w<-1$, however, $R_{\rm h}$ shrinks at late times and its maximum extent 
therefore occurred in our past. Under no circumstances will any null geodesic ever
reach us after having propagated to proper distances exceeding our cosmic horizon.

\acknowledgments
This research was partially supported by ONR grant N00014-09-C-0032
at the University of Arizona. I am grateful to Amherst College
for its support through a John Woodruff Simpson Lectureship. I
am also grateful to the anonymous referee for suggestions that
have improved the manuscript.

\end{document}